# Mapping Resilience and Inequality: A Spatial Analysis of Refugee Communities in Seattle

Rafi Khan, Bo Zhao

University of Washington

## Abstract

Seattle has long been a hub for refugee resettlement in the United States, resulting in diverse multicultural neighborhoods. This study provides an analysis of the refugee experience in Seattle and examines spatial patterns of housing cost burden, language isolation, religious infrastructure distribution, and crime incidence to understand challenges faced by refugees in these areas. We overlay these spatial findings with insights from literature on refugee integration, social vulnerability, and community safety. Results show that White Center and Rainier Valley experience high housing cost burdens and social vulnerability, reflecting significant financial strain and exposure to socioeconomic risks. Language barriers are pronounced, with a high concentration of residents not fluent in English, potentially limiting access to jobs and services. Access to religious and cultural institutions is limited in these neighborhoods, which may affect community cohesion and cultural preservation. Finally, while crime heatmaps indicate notable incidents citywide, evidence suggests refugees are more often victims than perpetrators and that the presence of refugee communities does not heighten crime rates. The paper concludes with recommendations emphasizing targeted housing support, language services, community space for cultural and religious practice, and inclusive public safety initiatives. This comprehensive analysis aims to inform policymakers, community organizations, and educators about the spatial and social dimensions of refugee experiences in Seattle, in a manner accessible to a high school–level academic audience.

## Introduction

Seattle is recognized as one of the most ethnically diverse and fastest-growing cities in the United States. It has a longstanding history as a destination for refugees and immigrants, with nearly 40% of the metro area's population being either foreign-born or having at least one foreign-born parent. Washington State consistently ranks among the top ten U.S. states for refugee resettlement, and between 2015 and 2019 alone it welcomed over 14,000 refugees. These newcomers hail from a wide range of countries, including Myanmar, the Democratic Republic of Congo, Somalia, Iraq, Ukraine, Afghanistan, and others. Many refugees who arrive in the Seattle area settle in communities with existing immigrant networks, such as White Center and Rainier Valley, which are known for their vibrant multicultural populations. White Center and Rainier Valley are two neighborhoods with significant refugee and immigrant presence. White Center is *"one of the most ethnically and income-diverse neighborhoods in Washington state, with a large immigrant and refugee population"*. Over 40% of White Center households are bilingual, and its poverty rate is nearly double the county average, indicating the socioeconomic challenges that many refugee families face. Rainier Valley, likewise, has been described as a *"gateway for immigration"* in Seattle's history and is often cited as one of the most diverse areas in the nation. In Rainier Valley's central zip code 98118, no single ethnic group holds a majority: as of the 2010 Census, about 33% of residents were Asian, 31% White, 26% Black, and 8% Hispanic, with over 59 languages spoken in the area. This extraordinary diversity reflects the multitude of refugee communities – from Southeast Asian to East African and beyond – that have made Rainier Valley their home over decades.

Despite their strengths in diversity and community bonds, these neighborhoods also concentrate on many of the *challenges* associated with refugee resettlement. Common issues include housing affordability (many refugee families struggle with high rents and cost of living), language barriers (limited English proficiency can hinder access to jobs, education, and services), and social vulnerability (refugees may lack economic stability or face health and transportation barriers). Additionally, access to cultural and religious institutions – such as mosques, temples, churches, and community centers that cater to refugee populations – may be limited, potentially affecting refugees' ability to maintain cultural practices and social support networks. Finally, perceptions of community safety and crime can influence refugee integration: refugees are sometimes stereotypically associated with crime or seen as contributing to insecurity, despite evidence to the contrary (Helfgott *et al.*, 2020). Understanding the reality of crime distribution in these areas and refugees' experiences with safety is crucial for dispelling myths and informing policy.

To explore these issues, this paper conducts a spatial analysis of White Center and Rainier Valley, using maps and data to visualize the refugee experience and its associated challenges. By overlaying these spatial data with findings from existing literature on refugees in Seattle and beyond, we aim to paint a comprehensive picture of how refugees in White Center and Rainier Valley are faring. The *central research questions* guiding this study are: (1) How do White Center and Rainier Valley compare to other Seattle areas in terms of housing affordability, language access, social vulnerability, and safety? (2) What do these spatial patterns reveal about the challenges and needs of refugee communities? (3) How do the experiences of refugees in these neighborhoods align with or diverge from broader research on refugee integration? By exploring these questions, this paper seeks to reveal the diversity of Seattle's refugee communities while also critically examining the structural challenges they face, thus offering recommendations for fostering more inclusive and resilient neighborhoods. Unlike prior qualitative studies on refugee experiences, this research integrates GIS spatial data with socio-cultural analysis to visualize inequities and resilience within Seattle's refugee neighborhoods.

## Literature Review

Seattle's refugee communities have been the subject of various studies spanning topics of cultural adaptation, social services, housing, mental health, and public policy. This literature review synthesizes key findings from prior research, providing context and depth to the spatial analysis. The review is organized around major themes relevant to the refugee experience: cultural integration, community support and services, housing and economic challenges, language and technology access, public policy environment, and safety and perceptions of crime.

Early research on Seattle's refugee populations highlighted the importance of cultural preservation in the adaptation process. For instance, Muecke (1987) examined *Lao refugees in Seattle* and their *"reconstruction of identity"* after resettlement. Even when displaced from their country of origin, Lao refugees found ways to practice and reinterpret their cultural traditions in the new context. Muecke noted that folklore and spiritual beliefs were adapted to Seattle's environment – famously, a traditional Lao ghost story was retold in a Seattle setting – as a means for refugees to make sense of their experiences and maintain continuity with their heritage. This study concluded that the ability of refugees to form communities with others of shared cultural background significantly helped them integrate more quickly and foster a sense of belonging. In other words, strong co-ethnic community ties and cultural expression can act as buffers against the disorienting effects of resettlement.

Beyond informal community networks, formal support structures are crucial for refugee integration. Kombassere (2013) investigated *refugee resettlement in Seattle* with a focus on the partnerships between community-based organizations (CBOs) and voluntary resettlement agencies (often faith-based or nonprofit "Volags"). In Seattle/King County, Volags handle initial refugee reception and placement, but longer-term integration often depends on local CBOs. Kombassere's analysis – drawing on reviews of agency reports and interviews with practitioners – found that collaboration between these entities was relatively weak and needed improvement. The central argument of this thesis was that stronger partnerships between resettlement agencies and grassroots community organizations lead to better outcomes for refugees, by combining the former's resources and the latter's cultural/community knowledge. CBOs such as the Ethiopian Community Mutual Association (ECMA, founded in the 1980s in Seattle) were highlighted as playing a key role in helping refugees navigate daily life and maintain cultural identity. Unlike larger agencies that might only assist during the initial months, community organizations often provide ongoing support (e.g. mentorship, language classes, job referrals) and build trust within refugee neighborhoods (Kombassere, 2013). This finding aligns with broader trends in refugee resettlement research that emphasize the value of "whole community" approaches – essentially, integration is most successful when local nonprofits, schools, health providers, and refugee community leaders are all working in concert.

Access to affordable housing and employment are persistent challenges for refugees rebuilding their lives. Kleit and Manzo (2013) found that refugees in redeveloped Seattle housing projects like Yesler Terrace and Rainier Vista faced high rents, limited affordable units, overcrowding, and discrimination after short-term aid ended. Language barriers and unfamiliar rental systems further hindered stable housing. Despite these constraints, many refugees preferred to live near co-ethnic communities and reliable transit, valuing proximity to familiar languages, culture, and jobs over larger or cheaper units. Liebow et al. (2004) examined Seattle's Jobs-Plus program, which served many refugee residents at Rainier Vista, and reported that 64% participated thanks to culturally responsive strategies—multilingual materials, bilingual caseworkers, and flexible scheduling. Their findings highlight that when housing and employment programs are tailored to refugees' linguistic and cultural needs, engagement and outcomes improve significantly.

Language proficiency is central to refugee integration, shaping access to education, employment, and social participation. In Seattle, many refugee adults arrive with limited English, and neighborhoods like White Center and Rainier Valley show the region's highest concentrations of language-isolated households. Dahya et al. (2020) found that limited English skills compound digital exclusion: many refugee women rely on smartphones with minimal data as their main internet access, struggle to use apps, and miss opportunities for online learning, job applications, and social services. Without language-accessible, hands-on training, technology can deepen inequality and isolation. Yet when libraries or community centers combined ESL and digital skills instruction, participation and confidence rose markedly, empowering women to manage daily tasks independently. Addressing these barriers requires culturally responsive strategies—bilingual education, community interpreters, and accessible tech-based learning—to promote long-term inclusion and self-sufficiency.

Seattle promotes itself as a "Welcoming City" for immigrants and refugees, but symbolic policies have not always produced tangible results. Davis (2020) found that Seattle's sanctuary resolutions—barring police from asking about immigration status and limiting cooperation with federal enforcement—offered reassurance but little material change. Many refugees remain unaware of these protections or distrust authorities due to past trauma, while inconsistent funding and implementation have limited real impact. Though Seattle's approach somewhat reduced fear compared to cities like Boston, it did not significantly improve policing relationships or service access. Davis concluded that sanctuary policies must pair symbolism with concrete measures such as legal aid, healthcare access, and community outreach. The creation of Seattle's Office of Immigrant and Refugee Affairs (OIRA) in 2012 reflects this intent, offering citizenship workshops, language programs, and civic engagement efforts—though its impact at the neighborhood level remains difficult to measure.

Safety and perceptions of crime are central to the refugee experience, influencing both well-being and community relations. Research shows a persistent gap between actual crime rates and residents' fear of crime, with refugees often caught in between—sometimes unfairly blamed, other times overly fearful due to past trauma. Helfgott et al. (2020) analyzed Seattle Police data and community surveys and found that fear often diverged from reality: some neighborhoods with low crime still reported high fear, shaped by factors such as homelessness visibility or racial bias. In areas with larger immigrant and minority populations, fear levels were often elevated despite modest crime rates, suggesting that prejudice and unfamiliarity distort perceptions. Importantly, refugees and immigrants were more often victims than perpetrators, particularly of bias-related incidents, and refugee communities generally exhibited low crime rates and strong internal social cohesion. These findings challenge misconceptions that diversity brings danger. Our analysis similarly explores whether White Center and Rainier Valley—home to many refugees—face elevated crime levels or stigma. Building trust through culturally informed community policing, multilingual communication, and liaison officers can help bridge these perception gaps and foster a greater sense of safety among refugees.

In summary, the literature depicts a complex but ultimately hopeful picture: Refugees in Seattle face significant hurdles – high living costs, language and digital divides, and gaps in formal support – yet they exhibit resilience through strong community bonds, cultural adaptation, and participation in programs when those are made accessible. White Center and Rainier Valley emerge from the literature as places of both concentrated disadvantage and concentrated community strength. These studies guide our expectations for the data: we expect to see spatial evidence of the challenges (e.g., maps showing high housing burdens and limited English proficiency in these areas) which correspond to the stories told in qualitative research. We also carry forward the insight that improving the refugee experience requires targeted, culturally competent interventions – a theme we will revisit in the Discussion, using both the map results and the literature to suggest pathways for action.

## Data & Methodology

This study employs a mixed-method approach to examine the refugee experience in Seattle. In this section, we describe the data sources, variables, and methods used to create the maps and interpret the results. The focus is on ensuring that the analysis is evidence-based and reproducible, even as it remains accessible to a high school audience. The primary geographic focus is on the neighborhoods of White Center and Rainier Valley. White Center is an unincorporated urban area in King County bordering the city of Seattle (just south of the Seattle city line in West Seattle/South Delridge). It roughly corresponds to the zip code 98146 and includes tracts known for large immigrant populations (e.g., around 15th Ave SW and Roxbury St). Rainier Valley is considered part of Seattle, encompassing several sub-neighborhoods such as Columbia City, Hillman City, Rainier Beach, and New Holly/Othello.

Data for this study were compiled from multiple reputable public sources to provide a multidimensional view of community conditions in Seattle's refugee-concentrated neighborhoods. Measures of housing cost burden and language isolation were drawn from the U.S. Census Bureau's American Community Survey (ACS, 5-year estimates). Housing burden was defined as the percentage of households spending more than 30% of their income on rent or mortgage plus utilities—a standard affordability threshold indicating financial strain. Language isolation identified households where no member aged 14 or older speaks English "very well," highlighting communities that may face barriers in education, employment, and access to public services. The Social Vulnerability Index (SVI), developed by the U.S. Centers for Disease Control and Prevention (2018), was used to assess broader socioeconomic and demographic risk factors. This composite index (ranging from 0 to 1) integrates data on poverty, crowded housing, lack of vehicle access, unemployment, and minority or language status to gauge each tract's capacity to respond to external stresses such as health crises or economic shocks. To evaluate cultural and social infrastructure, we compiled data on religious and community institutions from local directories and Google Maps searches, including churches, mosques, temples, synagogues, and other spaces known to serve immigrant or refugee groups. Each site was geocoded and plotted as a point on the map to identify clusters or gaps in access to such institutions. These data offer insight into the spatial distribution of places that foster belonging, cultural preservation, and informal support networks. For public safety, crime data were obtained from the Seattle Police Department's incident reports (January–September 2013). The dataset includes violent and property crimes and was analyzed using kernel density estimation to visualize "hotspots" of reported incidents. Although somewhat dated, these records capture enduring spatial patterns of crime concentration in Seattle and allow for comparative assessment of neighborhoods like White Center and Rainier Valley. Collectively, these datasets provide a comprehensive and spatially consistent foundation for examining the economic, linguistic, social, and safety dimensions of refugee community experiences.

Spatial data were projected onto a common map of Seattle. The maps were produced using ArcGIS and QGIS software. We cross-validated some of the ACS data with King County's Open Data portal to ensure accuracy. Margins of error in ACS for small-area estimates (e.g., in a single tract) mean that exact percentages should be interpreted with caution, but broad patterns (like one area having double the housing burden of another) are reliable. The SVI is a composite and already accounts for various correlated factors, but we use it primarily to give a single snapshot of "vulnerability." For the religious institutions data, one limitation is that it does not account for informal community worship (like home prayer groups) or very small congregations; it focuses on established centers. The crime data for 2013 provide a pre-2016 basemap of crime; we assume for our analysis that relative differences between neighborhoods (e.g., which areas have more crime) are roughly similar today, even if overall crime rates have fluctuated. By combining spatial analysis with socio-cultural research, our methodology provides a robust framework for analyzing the refugee experience in Seattle's neighborhoods. In the following Results section, we will present each map in Figure 1 along with explanations and then delve into a discussion that ties these findings back to the literature and the broader questions posed.

## Results

In this section, we present the key findings from the spatial analysis. Each sub-section corresponds to one of the thematic maps, supplemented by a figure and a narrative description. The figures illustrate how White Center and Rainier Valley compare with other Seattle neighborhoods regarding housing burden, language isolation, religious infrastructure, social vulnerability, and crime distribution. The results highlight significant spatial disparities that affect the refugee communities in these areas.

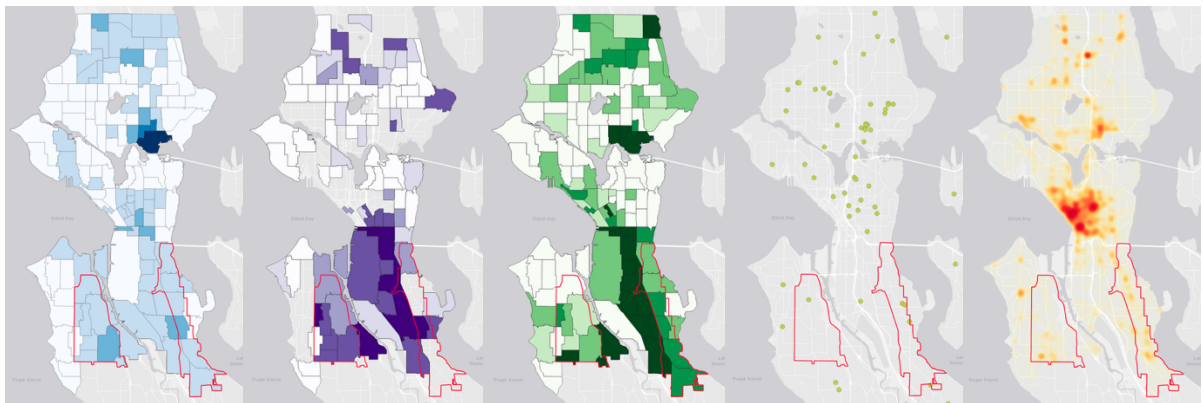

*Figure 1.* Map of Seattle: (a) Housing Cost Burden, (b) Social Vulnerability, (c) Language Isolation, (d) Religious Centers  and (e) Crime Concentration.

Figure 1(a) illustrates the share of household income spent on housing across Seattle, where darker blue shades represent heavier housing cost burdens. White Center and southern Rainier Valley appear among the darkest areas, with over half of households spending more than 30%—and many over 50%—of their income on housing, far higher than affluent neighborhoods north of downtown. This pattern reflects lower incomes and limited affordable housing options, forcing many refugee families—often large or multigenerational—to pool resources and live in crowded conditions. These findings align with Kleit & Manzo (2013), who observed refugees frequently paying far above affordable thresholds to remain near their communities. Such high rent-to-income ratios cause difficult trade-offs—reducing spending on essentials like food, healthcare, and education—and heighten displacement risk if rents continue to rise. The data highlight the need for targeted rental assistance, affordable housing development, and income supports in these areas. Promising initiatives like the White Center Community HUB, which will include 76 affordable apartments, represent meaningful progress toward closing this gap.

The heavy housing burden in White Center and Rainier Valley directly contributes to their broader social vulnerability. Financially strained households have less capacity to handle emergencies or invest in opportunities such as education or small businesses. As shown in Figure 1(b), the Social Vulnerability Index (SVI) map (purple shading) places White Center among the county's most vulnerable areas, with a continuous stretch of dark purple indicating overlapping challenges—high poverty, unemployment, limited vehicle access, and crowded housing. Rainier Valley shows a similar pattern, with dark purple hotspots around New Holly and Rainier Beach and moderate vulnerability near Beacon Hill. Both neighborhoods rank far higher in vulnerability than most of North Seattle, which appears light purple or white on the map. These findings confirm that refugee-concentrated areas face compounded socioeconomic and infrastructural risks. The figure notes that such conditions "exemplify the difficulties faced by refugee and immigrant residents" and call for policy adjustments. In practical terms, this means more targeted interventions—multilingual disaster preparedness, accessible social services, and economic supports—are needed. Consistent with Kombassere (2013), the SVI results quantitatively affirm that existing systems have not adequately met the needs of these communities, reinforcing the urgency of strengthening local support networks.

Figure 1(c) maps the distribution of households with limited English proficiency across Seattle, shaded in green. Darker areas indicate higher concentrations of linguistically isolated households, with White Center and Rainier Valley showing the darkest green zones—some tracts exceeding 30–40% of households where no adult speaks English well. By contrast, most of North Seattle and downtown appear pale green or nearly white (under 10%), revealing a sharp spatial disparity. These data confirm that refugees in these neighborhoods are far more likely to live in non-English-speaking enclaves. While such clusters provide mutual support, they can also reinforce social isolation and limit access to jobs, education, and public services. The map supports what service providers consistently report: many residents require translation for routine tasks like healthcare visits or parent-teacher meetings. This aligns with Dahya et al. (2020), who found that language barriers often compound digital and economic exclusion. Schools in Rainier Valley, where dozens of home languages are spoken, face especially high demand for ELL programs and bilingual staff. At the same time, these neighborhoods' linguistic diversity has fostered strong community responses—ethnic centers offering ESL classes, and libraries hosting language-exchange meetups. The data suggest that White Center and Rainier Valley should be top priorities for expanding interpreters, multilingual materials, and culturally tailored digital literacy training to transform linguistic diversity from a barrier into an asset for inclusion.

Figure 1(d) maps the distribution of religious and spiritual centers across Seattle, with a focus on those serving immigrant and refugee communities. Each dot represents a church, mosque, temple, or similar institution, and the outlined areas highlight White Center and Rainier Valley. The map reveals a clear disparity: White Center has only one major religious site within its boundary, while Rainier Valley shows a few scattered places of worship—still far fewer than areas like North Seattle or the International District, where institutions cluster densely. This shortage is significant because faith-based spaces often serve as more than sites of worship— they are hubs for social support, cultural identity, and information exchange. As Kombassere (2013) notes, community well-being depends not only on material stability but also on cultural infrastructure. The lack of accessible religious centers in these neighborhoods means many refugees must travel long distances or rely on informal prayer groups in apartments or storefronts, which may lack permanence and capacity. For elderly or mobility-limited residents, this distance can lead to isolation and a loss of routine spiritual practice. The limited access to worship spaces impacts the ability for refugee populations to practice their religion and spiritual lives and weakens community bonds. Addressing this gap requires intentional community planning—city agencies and nonprofits can help congregations secure property, provide grants, or adapt existing public buildings for multicultural use. While recent initiatives like the expansion of the Somali Community Center in Rainier Valley show progress, the overall distribution of religious and cultural spaces remains uneven, highlighting the ongoing need for inclusive infrastructure that supports spiritual and communal life.

Figure 1(e) presents a crime heat map of Seattle based on 2013 police-reported data. Areas shaded in deeper red indicate a higher concentration of reported incidents, while yellow or lighter tones represent fewer reports. As expected, downtown Seattle and its surrounding neighborhoods (such as Capitol Hill and Belltown) appear as bright red "hotspots," reflecting the high volume of incidents typical in dense urban and nightlife areas. In contrast, White Center and Rainier Valley do not display extreme crime intensity; most of these areas appear orange or light red, indicating moderate concentrations of reported crime. A few orange-tinted zones are visible along Rainier Avenue S and near some transit centers, but overall, their intensity is far lower than the city center. White Center shows a similar pattern—moderate density, but not a high-crime hotspot.

It is important to emphasize that this map is derived from police reports, not judicial or conviction data. There is often a gap between reported and actual crime: some cases are reclassified, dropped, or never reach prosecution. Thus, the heat map reflects where incidents were recorded by police, not necessarily where legally verified crimes occurred. Moreover, the map does not distinguish between crime types or severity—a minor property theft and a violent assault carry equal weight—making this a rough estimate rather than a precise measure of risk. Another structural issue to acknowledge is over policing. In lower-income and racially diverse neighborhoods, police patrols and surveillance tend to be more frequent, which can inflate report counts even when actual crime rates are not higher. Research shows that minor infractions or public disturbances are more likely to be recorded in minority communities, amplifying the perception of high crime. In this sense, the map reflects not only patterns of criminal activity but also the uneven distribution of law enforcement attention across Seattle.

White Center and Rainier Valley do not exhibit higher crime levels than other parts of the city. In fact, some wealthier northern and central neighborhoods report comparable or higher per-capita property crime rates (e.g., vehicle thefts). This finding supports Helfgott et al. (2020) and others, who concluded that immigrants and refugees are not drivers of crime. On the contrary, refugees are often victims rather than perpetrators, and the data confirm that "refugees have no role in increasing Seattle's crime rates." This insight is essential for public perception. White Center and Rainier Valley are sometimes stigmatized as "unsafe" because of their diversity and lower incomes, but data do not support this narrative. The real issues lie in resource inequities and uneven policing. Refugees may hesitate to report crimes due to language barriers, prior trauma, or mistrust of law enforcement, while heavy patrol presence can make them feel over-scrutinized. Community surveys also reveal that Rainier Valley residents often report higher fear of crime than actual risk levels (as discussed in the literature review). Future safety strategies should therefore focus not only on reducing crime but also on rebuilding trust between communities and police. Initiatives such as multilingual community policing, diverse officer recruitment, improved public lighting, and youth engagement programs (e.g., Rainier Beach Action Coalition's late-night sports activities) can enhance both real and perceived safety. Ultimately, this heat map should be viewed as a spatial projection of policing data, not a definitive measure of criminal reality. It reflects a complex social landscape where refugee communities—despite facing structural inequality and risks of overpolicing—maintain strong cohesion and low crime rates. Policymakers should therefore prioritize equitable policing and community trust-building as key components of inclusive public safety.

To summarize the results: the maps collectively depict high housing cost burdens, high social vulnerability, significant language barriers, limited formal community infrastructure, yet only moderate crime levels in White Center and Rainier Valley. These spatial patterns align with what one would expect in neighborhoods that are home to many newly arrived refugees and other immigrants. The data paint a picture of communities under economic and social strain but also dispel myths (for example, about crime). In the next section, we delve into what these results mean – interpreting how these conditions affect refugee lives on a daily basis, and discussing how they correlate with the academic and policy insights from the literature.

## Discussion

Together, the spatial patterns of housing burden, language isolation, and limited community infrastructure converge to form a multi-layered geography of inequality that both constrains and reveals refugee resilience. The maps show that residents of White Center and Rainier Valley face disproportionately high housing costs, echoing findings by Kleit and Manzo (2013) that refugee families struggle to afford Seattle's housing. Many arrive with limited resources and rely briefly on rental aid before confronting the harsh tradeoff between low wages and high rents. In some tracts, over half of households spend more than 30%—often over 50%—of income on housing, forcing families to share crowded units and sacrifice essentials like food or healthcare. The Social Vulnerability Index further indicates that these neighborhoods concentrate multiple socioeconomic risks, making economic instability one of the most pressing challenges for refugees. Suggested responses include expanding affordable housing through subsidies or nonprofit development, offering rental and eviction assistance, and investing in job training and small business programs. While Seattle's housing initiatives in Rainier Valley show progress, the gap between demand and available support remains wide, leaving many refugee families under continuing financial strain. Despite heavy housing burdens, refugee communities often stabilize rather than strain neighborhoods. In Rainier Valley, steady in-migration and small business activity have prevented the disinvestment seen in other U.S. cities. Property values in these diverse areas have risen gradually, challenging the misconception that refugees lower neighborhood value. As renters, consumers, and entrepreneurs, refugees strengthen local economies—Somali bakeries and Vietnamese markets, for instance, have revived South Seattle's commercial corridors. Yet gentrification and rising rents now threaten to displace these very communities. Policies such as community land trusts, stronger tenant protections, and targeted affordable housing are essential to preserve their presence. As Kleit and Manzo (2013) note, relocation to distant suburbs can sever refugees from public transit and support networks, underscoring the need to maintain accessible, affordable housing within the city to sustain both community vitality and stability.

White Center and Rainier Valley have some of the highest concentrations of residents with limited English proficiency, affecting employment, education, and daily communication. As Dahya et al. (2020) noted, language barriers often compound other challenges such as technology use and healthcare access. Programs like Seattle's *Ready to Work* have begun addressing these gaps by combining ESL and job training, but demand far exceeds capacity. Local organizations—including the Somali Youth & Family Club and Refugee Women's Alliance—are crucial in providing interpretation and culturally tailored support, yet need greater funding and partnership. At the same time, multilingualism is a major community asset: over 60 languages are spoken in Rainier Valley, reflecting strong cultural and linguistic capital. Expanding bilingual education, heritage language programs, and multilingual civic outreach can transform linguistic diversity from a barrier into a bridge for inclusion, family connection, and long-term integration.

White Center and Rainier Valley have among the highest concentrations of residents with limited English proficiency, a reality that affects nearly every part of daily life—from employment and healthcare access to communication with schools and public institutions. As Dahya et al. (2020) observed, language barriers often compound digital and social exclusion, leaving many refugee adults—especially women—unable to fully participate in civic and economic life. Programs such as Seattle's Ready to Work initiative, which integrates ESL instruction with job skills training in immigrant-dense neighborhoods, represent an important step toward inclusion. However, our findings indicate that far more resources are needed. Schools serving these areas would benefit from additional bilingual teachers and family engagement specialists to bridge communication gaps with parents, while libraries and community centers could expand evening and childcare-supported ESL programs. Community-based organizations such as the Somali Youth & Family Club in White Center and the Refugee Women's Alliance (ReWA) in Rainier Valley already play a pivotal role by offering interpretation, technology assistance, and culturally responsive workshops, yet their reach remains limited by funding constraints. Supporting these grassroots groups through city partnerships or grant programs is one of the most direct ways to reach linguistically isolated households. At the same time, the linguistic diversity of these neighborhoods—over 60 languages spoken in Rainier Valley alone—should be recognized as an asset rather than a deficit. Bilingual and heritage-language education has been shown to strengthen academic outcomes and family cohesion, and Seattle Public Schools' existing dual-language programs could be expanded to include languages common among refugees, such as Somali, Amharic, or Arabic. Broader civic inclusion efforts, like the city's Ethnic Media Program and multilingual voter outreach, also deserve continued investment. Collectively, these efforts would not only reduce isolation but transform linguistic diversity into a foundation for equity, participation, and cultural vitality in Seattle's refugee communities.

The shortage of religious and community centers in White Center and Rainier Valley highlights a critical gap in the social fabric of refugee life. While many groups self-organize—renting small rooms for worship or cultural events—these temporary arrangements cannot replace dedicated, permanent spaces for gathering and support. As noted by Muecke (1987) and Kombassere (2013), communal spaces are essential for preserving identity, fostering belonging, and maintaining emotional well-being. Investing in such infrastructure should be viewed as equally important as providing housing or jobs. Support could include grants or technical assistance to help community-based nonprofits acquire facilities, or reforms to zoning and permitting that make it easier for religious organizations to establish centers. Promising initiatives like the forthcoming White Center Community HUB, designed as a multi-service complex offering healthcare and community space, reflect what residents themselves have called for—safe, accessible, and culturally inclusive places to gather. Rainier Valley's Southeast Seattle Community Center serves a similar function but remains insufficient given the area's diversity and size. Our findings also reveal that these neighborhoods are not divided by religion or ethnicity; a single mosque or church may serve attendees from several national backgrounds. This diversity, while enriching, also presents logistical challenges for language and programming. Therefore, developing multipurpose, interethnic community centers—spaces that unite rather than divide—offers the most effective way to strengthen cohesion and provide equitable access to resources across refugee communities.

Our crime analysis shows that White Center and Rainier Valley are not high-crime areas compared to Seattle's commercial or nightlife districts, yet residents often perceive greater danger than data supports (Helfgott et al., 2020). This perception gap—shaped by trauma, language barriers, and occasional bias incidents—requires both communication and trust-building. Public messaging should clarify that refugee communities do not drive crime; in fact, many immigrant neighborhoods report lower rates due to strong family and social networks. At the same time, addressing refugees' fear of crime is essential, especially for those who may distrust law enforcement based on past experiences. Initiatives like multilingual safety meetings, diverse police recruitment, and Refugee Women's Alliance workshops that explain legal rights can improve relationships between refugees and authorities. Because refugees are often victims rather than perpetrators, accessible and culturally competent victim services remain crucial. Organizations such as API Chaya and the Crime Victim Service Center provide important multilingual assistance but need broader support. Safety also intersects with infrastructure—better lighting, well-maintained parks, and youth engagement programs (like Rainier Beach Action Coalition's late-night sports events) all contribute to real and perceived security. Ultimately, combining community-led initiatives with inclusive policing offers the best path to fostering safety and trust in refugee neighborhoods.

A key theme emerging from this study is the strong interconnection among housing, language, employment, safety, and culture. A refugee family with limited English may only access low-wage jobs, leading to crowded or less safe housing and limited mobility, which in turn restricts participation in community or religious life. These challenges reinforce one another, but targeted improvements in one area—such as access to affordable housing or language education—can create ripple effects across others. The literature (Kombassere, 2013; Dahya et al., 2020) similarly emphasizes the need for holistic, place-based approaches that integrate housing, education, and economic opportunity within the same communities.

Compared with other U.S. cities that host refugee clusters, Seattle's high cost of living and technology-driven economy intensify these pressures, widening the gap for refugees with limited or non-transferable skills. This context calls for locally adapted solutions—such as combining tech and language training—to help refugees access new industries. At the same time, Seattle's strong philanthropic and corporate sectors, including Microsoft, Amazon, and the Gates Foundation, present major opportunities to strengthen refugee initiatives. Leveraging these resources alongside public investment could build a more inclusive, coordinated framework for integration and resilience.

While much of the discussion has centered on needs and challenges, it is also crucial to acknowledge the resilience and assets of these refugee communities. White Center and Rainier Valley are places of entrepreneurial spirit – family-owned restaurants, markets, and small businesses line the streets (from Halal butchers to ethnic clothing shops). These businesses serve as economic engines and cultural ambassadors, and they often hire within the community, providing jobs and training. Socially, the communities have self-formed networks like savings groups, faith groups, and advocacy coalitions. For example, groups of mothers in Rainier Valley formed a cooperative childcare to mind each other's kids in a culturally appropriate way so they could attend ESL classes. During the COVID-19 pandemic, mutual aid networks sprang up in these areas, with younger bilingual refugees helping elders navigate vaccine appointments and delivering food. Such stories exemplify strong social capital. From an academic perspective, this aligns with theories of social resilience – despite high "vulnerability" scores in metrics like SVI, the on-the-ground reality might be buffered by these strong informal support systems that numbers don't fully capture.

Effective interventions must be designed with refugee communities rather than for them. As Kombassere emphasizes, community-based organizations are best positioned to lead such efforts, with public agencies and nonprofits serving as partners and supporters. White Center's new Community HUB—born from resident-led summits—is a model example of how locally driven planning can generate practical, trusted solutions. Building on this approach, several policy directions emerge from our findings. First, Seattle should prioritize affordable housing in White Center and Rainier Valley, expanding development, rental assistance, and community land trusts to prevent displacement and ensure long-term affordability. Second, targeted language and employment initiatives are essential: more ESL and vocational programs within the neighborhoods, on-site workplace English classes, and bilingual tutoring for refugee youth can close the opportunity gap. Equally critical is investment in cultural and religious infrastructure. Grants, zoning support, and adaptive reuse of public buildings could enable community groups to establish multi-purpose spaces for worship, education, and gathering—anchors of identity and inclusion. Public safety strategies must also evolve toward partnership and trust: recruiting multilingual officers, supporting youth engagement programs, and ensuring victims have access to culturally sensitive reporting systems. Beyond specific sectors, a holistic, place-based development framework is needed. Initiatives like Seattle's Equitable Development Initiative (EDI) demonstrate how coordinated investment in housing, public health, and cultural capacity can build lasting community resilience. Finally, combating misconceptions about refugee neighborhoods through education, journalism, and data transparency remains essential. Sharing evidence that these communities are safe, vibrant, and economically active can counter stigma and foster empathy. By aligning policy, philanthropy, and grassroots leadership, Seattle can strengthen these neighborhoods as inclusive, thriving models of urban refugee integration.

## Conclusion

Seattle's neighborhoods of White Center and Rainier Valley vividly illustrate the realities of contemporary refugee life in an urban context. Through spatial mapping and socio-cultural analysis, this study revealed patterns of resilience alongside structural inequities. Refugees in these areas have rebuilt their lives, opened businesses, and enriched Seattle's multicultural identity, yet continue to face steep housing costs, language barriers, and economic precarity. High housing burdens and elevated social vulnerability indexes highlight persistent affordability and access gaps, showing that existing support systems remain insufficient.hao'ya Because refugee populations are geographically concentrated, interventions should be place-based— expanding affordable housing, reliable transit, and community health services directly within these neighborhoods. Programs such as the White Center Community HUB and the Equitable Development Initiative in Rainier Valley exemplify promising, community-driven models that merit continued investment and scaling.

Social infrastructure is equally vital. Schools, religious centers, and public gathering spaces provide not only essential services but also the foundation for belonging, cultural preservation, and mutual aid. Their absence can isolate refugees, while their presence strengthens community cohesion and accelerates integration. Our findings also counter persistent misconceptions linking refugees to crime. Both spatial data and prior research indicate that refugees are often victims rather than perpetrators, and that their neighborhoods exhibit strong social cohesion and moderate crime levels. Culturally competent policing, translation access, and community liaisons can further enhance safety and trust.

While the results are robust, several limitations should be noted. The crime dataset used dates from 2013, and shifts in policing, gentrification, and rising gun violence since 2020 may have altered neighborhood patterns. The inventory of religious centers may underrepresent small or newly established sites, and tract-level indicators such as the SVI and housing burden can mask variation within neighborhoods. Additionally, health outcomes—such as trauma or barriers to medical access—were beyond the scope of this study but remain critical for understanding refugee well-being. Despite these constraints, the convergence between spatial findings and existing qualitative research strengthens confidence in the validity of our conclusions.

Future research should build on this foundation by examining educational outcomes, long-term mobility patterns, and the integration of newer refugee groups such as Syrian, Afghan, and Ukrainian arrivals. Longitudinal GIS analyses could also track how ongoing community investments reshape vulnerability and opportunity in these neighborhoods. Ultimately, addressing the intertwined challenges of housing, language, culture, and safety demands coordinated, cross-sector collaboration among city agencies, schools, healthcare providers, and refugee-led organizations. Seattle's Immigrant and Refugee Commission offers a model of inclusive governance grounded in lived experience. Investing in refugee communities—through affordable housing, job and language programs, and inclusive public spaces—is an investment in Seattle's collective future. White Center and Rainier Valley stand as testaments to both the hardships and hopes of resettlement, reminding us that a truly welcoming city not only receives newcomers but empowers them to thrive—making diversity a lasting source of strength for the entire community.